\documentclass[twocolumn,showpacs,prl,preprintnumbers,amsmath,amssymb]{revtex4}
\usepackage{bm}
\usepackage{graphicx}
\usepackage{color}
\begin{document}
\title{Terahertz radiation due to random grating coupled surface plasmon polaritons
    }
\author{T.~V.~Shubina$^1$} 
\author{N.~A.~Gippius$^{2,3}$}
\author{V.~A.~Shalygin$^4$}
\author{A.~V.~Andrianov$^1$}
\author{S.~V.~Ivanov$^1$}

\affiliation{$^1$Ioffe Physico-Technical Institute, RAS, St. Petersburg 194021, Russia}
\affiliation{$^2$LASMEA, UMR 6602, Universite Blaise Pascal, 63177 Aubiere Cedex, France}
\affiliation{$^3$General Physics Institute RAS, Moscow 119991, Russia}
\affiliation{$^4$St. Petersburg State Polytechnical University, St. Petersburg 195251, Russia}

\begin{abstract}

We report on terahertz (THz) radiation under electrical pumping from a degenerate semiconductor possessing an electron accumulation layer. In InN, the random grating formed by topographical defects provides high-efficiency coupling of surface plasmon polaritons supported by the accumulation layer to the THz emission. The principal emission band occupies the 2-6 THz spectral range. We establish a link between the shape of emission spectra and the structural factor of the random grating and show that the change of slope of power dependencies is characteristic for temperature-dependent plasmonic mechanisms.  The super-linear rise of a THz emission intensity on applied electric power provides advantage of such materials in emission yield.

\end{abstract}

\maketitle

The surface plasma waves confined to the interface between a conductor (metal or degenerate semiconductor) and a dielectric attract a lot of attention for terahertz (THz) frequency range application \cite{Dyakonov,Hopfel}. The characteristic frequency   of a surface plasmon at the sharp interface of the semi-infinite conductor with vacuum reads as
\begin{equation}\label{eq1}
\omega_s^{2}=4{\pi}{N}{e}^{2}/m^{\ast}({\varepsilon}+1),
\end{equation}
where $N$, $e$, and $m^{\ast}$  are the bulk concentration, charge, and effective mass of carriers, respectively; $\varepsilon$  is the dielectric permittivity of the conductor. This frequency is in the visible range for the metals, being in the far infrared and even in the terahertz ranges for the degenerate semiconductors \cite{Hopfel,Allen,Rivas,Anderson}. The distinct difference between plasmon performances in these two types of materials is that the strong interband absorption in the visible range hampers collective electron excitations in the metals, while in the semiconductors the plasmon frequency is below an absorption edge. Thus, long-range surface plasmon polariton modes can be supported.

For strictly two-dimensional plasmons, the frequency depends on the plasmon wave vector $k$ as ${\omega}(k)\propto\sqrt{k}$ \cite{Ritchie}. (This relation is more complicated for plasmons in the films of a finite thickness \cite{Economou,Burke}.) It determines why the surface plasmons cannot radiate to ambient directly. Their wave vector is always higher than that of light $k_0=\omega/c$  at the same frequency. Coupling of the plasmon with an electromagnetic wave occurs by means of the momentum exchange either at the disturbances induced by plasma wave instability \cite{Dyakonov} or at the periodical modulations of surface profile \cite{Hopfel,Allen,Rivas}. For ${k}\gg{k_{0}}$, the radiating states have the momentum ${k}\approx{2}{\pi}{m}/{a}$  (${m}=1,2,\ldots$),  where $a$ is a grating period. In the case of the ideal grating, the $k$ intervals where the emission yield takes place are inevitably narrow and the integral emission intensity is not sufficient sometimes for reliable registration by available detectors. To circumvent this obstacle, more complicated (multiple-sections) gratings have been proposed \cite{Meziani,Wilkinson,Otsuji}.

 \begin{figure} [t]
\includegraphics{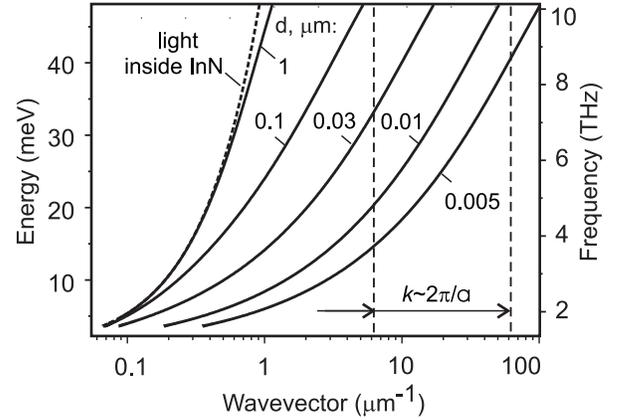}
  \caption{ \label{f1} Dispersion dependencies (semi-logarithmic scale) calculated for the different thicknesses $d$ of plasmon supporting layers shown together with the light wave dispersion inside InN. The crossing of vertical dashed lines with plasmon dispersion curves indicates the $d$ thickness range suitable for the first order coupling (${k}\sim{2}{\pi}/{a}$) for ${a}={0.1}{\div}1$ $\mu{m}$.
   }
\end{figure}

In this letter, we report on THz radiation  from a degenerate semiconductor possessing a carrier accumulation layer which can support surface plasmon polaritons. We demonstrate advantages of the random grating to provide effective coupling between these plasmons and the THz emission in a wide spectral region extended for several THz. The genesis of the THz emission is considered to establish the interplay of the distribution of grating wave vectors and a spectra shape. It is elucidated how the thermal population of plasmonic states controls the spectra and power dependencies.

The study has been done using InN epilayers which have an intrinsic electron concentration of $10^{18}-10^{19}$ cm$^{-3}$, that guarantees  ${\omega}_{s}$ within a terahertz range. InN has a tendency to the formation of various structural defects, such as pores, pits, and trenches, frequently decorated by metallic nanoparticles. The average distance between these defects is about the grain size ($\sim$1 $\mu$m); therefore, a certain periodicity exists in their topographic position. The 1-3 $\mu$m thick InN layers were grown by plasma-assisted molecular beam epitaxy on c-sapphire atop of GaN buffers at substrate temperature $T_{s}=(450-550)^{\circ}$C. For the THz electroluminescence measurements, contacts are formed on the top surface of samples. The THz spectra were measured at 8 K using a step-scan Fourier spectrometer with excitation by the series of packets of 15 V rectangular pulses of 10 $\mu$s duration (71.5 Hz repetition rate). The dependencies of integral THz emission intensity on the electric field were measured at 4.2 K using 2-$\mu$s-long pulses by a Ge:Ga photodetector with 2.4-5 THz spectral sensitivity range. Other experimental details are described elsewhere \cite{Shubina}.

Calculations based on numerical solving of the Maxwell equations for multiple layers structures have been used to model the full set of plasmonic and waveguide modes supported in the studied samples. They have shown that the  range of plasmonic layer thicknesses corresponding to the experimentally observed emission frequencies of several THz is rather narrow. For the sake of demonstration, we present in Fig. 1 the plasmon dispersion curves calculated using the three-layers model as reported previously \cite{Burke,Shubina}. In these calculations, the metal-like layer described by the Drude model (${N}={10}^{19}$ cm$^{-3}$, ${m}^{\ast}=0.13{m}_{0}$, damping parameter of 1 meV) is placed between the vacuum and the rest part of the InN epilayer, which is considered as a dielectric. The static permittivity of InN was assumed equal to 14 everywhere. The results of the calculations explicitly display that to provide the coupling by the typical InN defects, the layer thickness $d$ should be small enough, $\leq0.1$ $\mu$m. This estimation is consistent with similar limitation on the $d$ value needed to support the long propagating surface modes \cite{Economou,Burke}.  A good candidate for such a thin  plasmon-supporting layer is an electron accumulation layer on a top of a semiconductor film \cite{Kao}. The presence of the layer in InN has been confirmed by the spectroscopy of electron energy losses induced by the surface plasmons \cite{Mahboob}.

Here, we demonstrate other observation related presumably to the surface plasmon-polaritons, which is obtained by scanning electron microscopy (SEM). Before to its description, it is worth reminding that the electron beam excitation is a very efficient way  to create the surface plasmons \cite{Abajo}, and that this excitation is followed by the generation of secondary electrons (SEs) due to non-radiative decay of plasmons into single electron excitations \cite{Warmack,Lecante}.
The SEM studies were done at low temperature to minimize the plasmon damping. The electron beam impinges normally to a cleaved facet of a layer. This configuration corresponds to the so-called aloof scanning mode, when the electrons moving along nontouching trajectory above the sample surface can stimulate the surface plasmons in conductors \cite{Warmack,Lecante}.

The SEM images of the InN epilayers exhibit intricate features - clouds above the sample surface, which have a specific semispherical shape (Fig. 2). Their sizes increase with exposure to the electron beam. With slight declining a sample, it was possible to resolve discrete structure in their base, whose step is approximately equal to the average distance between the topographic defects. The plasmon excitation in the aloof mode implies that the defects scatter the electrons with consecutive charge accumulation. We suppose that the secondary electrons display an electric field around the charged defects in the same way as the iron filings depict the magnetic field lines. Since SEs can escape from rather small depth of several nanometers only, the registration of such clouds confirms implicitly the propagation of the plasma waves in a very thin electron accumulation layer on a top of a semiconductor.
\begin{figure} [t]
\includegraphics{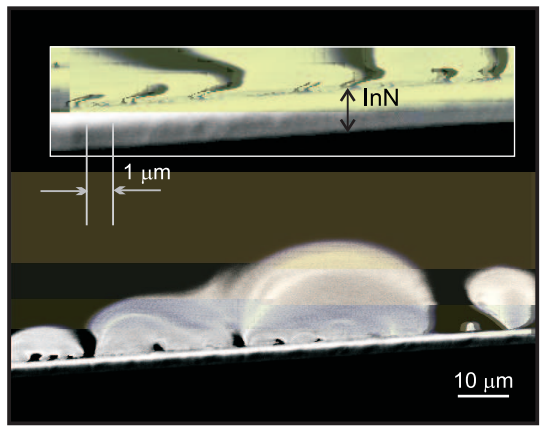}
  \caption{ \label{f1} Facet SEM image of an InN epilayer recorded at 5 K using a SE2 detector of a SEM analytical microscope (20 keV electron beam). The inset presents the image taken with a sample detuned slightly to show discretion in a cloud base.
   }
\end{figure}

To consider the genesis the THz emission from the InN layer under electrical pumping, we assume first that each harmonic of the random grating translates the corresponding plasmons into a light cone at the frequency governed by both the plasmon dispersion and the particular wave vector. To find the structure factor of the random grating, ${S}({k})$, the Fourier transformation of a SEM image was performed [Fig. 3 (a, b)]. Then, it is quite reasonable to suggest that the intensity of the radiation at certain energy is proportional to the amplitude of respective harmonic. The influence of the plasmon dispersion, calculated as described above, is taken into account by multiplying on the plasmon density of states in this point. Such an approach permits us to simulate directly an emission spectrum, neglecting higher order scattering. Figure 3 (c) demonstrates the good agreement between the computed and experimental spectra, in spite of a rather complicated shape provided by the random grating.
\begin{figure} [t]
\includegraphics{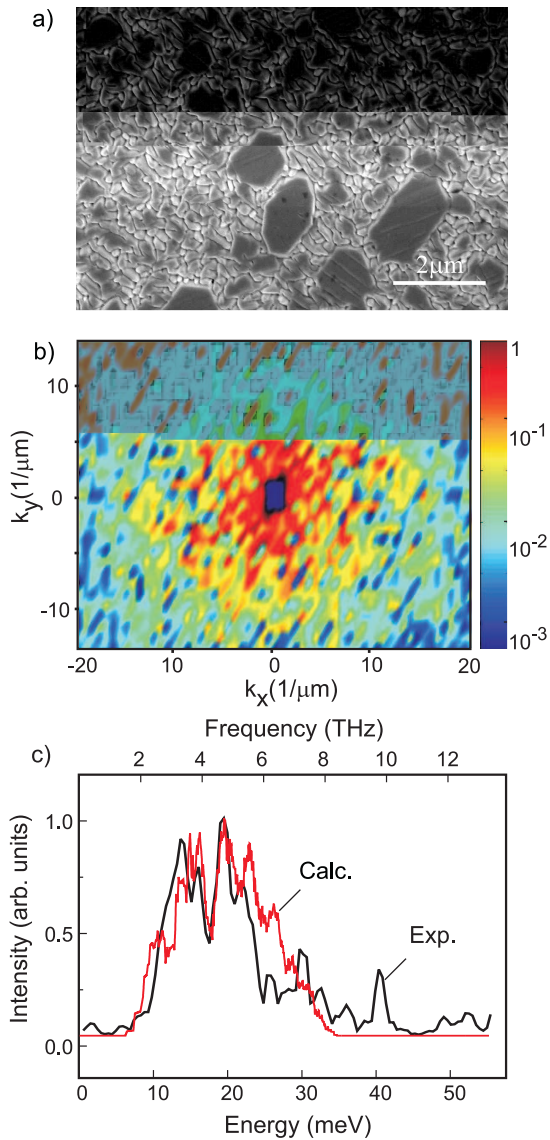}
  \caption{ \label{f1} (color online) Simulation of THz emission spectra. (a) SEM image of a surface of an InN epilayer (sample A). (b) Image of the structure factor of random grating obtained by the Fourier transformation of the SEM image. Color scale gives the harmonic amplitudes. Dark spot in the center is due to removing the unnecessary harmonics of lowest orders. (c) Comparison of experimental and calculated ($d=7$ nm, other parameters are given in the text) spectra.
   }
\end{figure}

In accord with the energy and momentum conservation requirements, the higher energy electrons populate plasmonic states with lower k-vectors \cite{Popov}. In our calculations, we ignore such a nonequilibrium plasmon population that provides, likely, some underestimation of the intensity of the lower-energy peaks in the calculated spectrum as compared with the experimental one.  The other difference in the spectra concerns the higher energy cut-off of the main emission band, which is at higher energy in the simulated spectrum. The difference arises out disregarding the finite population of the plasmonic states. At electrical pumping, the filling level is controlled by the electron temperature $T$, which can markedly exceed the lattice temperature \cite{Hopfel,Vitlina}. Its value is roughly proportional to the electrical power ${P}_{el}$. Our spectral measurements were done using a spared regime, i.e. with the small ${P}_{el}$; thus, the actual filling level was lower than the energy corresponding to the largest ${k}^{max}$  in the grating wave vector distribution.

The filling effect influences also the dependencies of emission intensity on electric power, namely, their characteristic slopes. To describe that, we suppose the plasmons to be thermalized with temperature $T$, e.g., via the electron scattering by phonons, impurities, and other defects \cite{Vitlina,Richter}. The emitting power ${W}({T})$ can be estimated by integration over all plasmon states with the energy ${E}({k})$, occupation of these states, and the structural factor of the grating ${S}({k})$:
\begin{equation}\label{eq2}
{W}({T})=\int{k}{d}{k}{S}({k}){E}({k})/({\exp}({E}({k})/{T})-1).
\end{equation}
Here, the temperature $T$ is given in the energy units. Two following limiting cases are considered: i) homogeneous emission from all possible plasmon states assuming ${S}({k})$=const; ii) emission from the grating-coupled  states with ${S}({k})$ derived from the SEM image (see Fig. 3).

The performed calculations show that the mechanism based on the uniform scattering dominates when  ${T}\leq{E}({k}^{min})$, i.e. when the states corresponding to the minimal grating wave vector are not populated yet [Fig. 4 (a)]. Consequently, the slope in the resulting emission power dependency must have the exponent  ${\beta}\sim{5}$. (This ${T}^{5}$ law appears due to the superposition of the Planck distribution function and the plasmon dispersion.) With the temperature rise, the mechanism of the emission changes - now the most of the emission comes from the plasmons interacting with the random grating. When the temperature approaches the higher energy cut-off  ${T}\sim{E}({k}^{max})$, the ${\beta}\sim{1.3}$  is expected. The predicted slopes and their change do be observed in the experimental dependencies of the radiated intensity on the electric power [Fig. 4 (b)]. Note that the distinctiveness of the kink and the corresponding ${P}_{el}$  value are specific for each sample, being dependent on the grating pattern and plasmon dispersion.

\begin{figure} [t]
\includegraphics{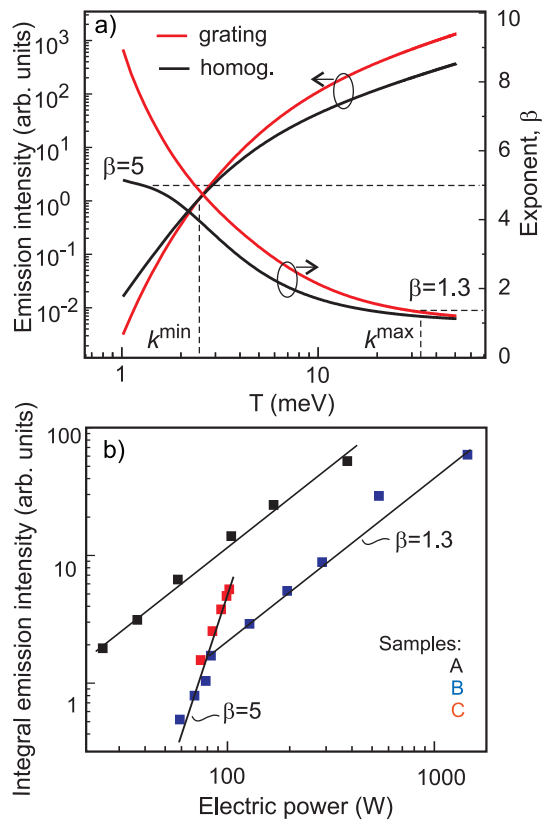}
  \caption{ \label{f1} (color online) (a) Dependencies of the emission intensity on temperature (given in energy units) calculated for the homogeneous filling of plasmon states and for the filling of grating-coupled states. The vertical lines mark the interval corresponding to the occupation of the grating-coupled states in the sample A. At low temperature, when the emission provided by the homogeneous filling dominates, the exponent $\beta$ is about 5; at the higher-energy boundary, it drops to $\sim$1.3. (b) Experimental dependencies of the emission intensity on the electrical power of representative samples exhibiting the similar slopes and the kink between them. The lines ${\beta}=5$ and ${\beta}=1.3$ are given to guide the eye.
   }
\end{figure}
For the sake of clarity, the presented consideration is markedly simplified. Nevertheless, it displays the basic features of the plasmonic emission from nanostructured degenerate semiconductors. In particular, the super-linear rise of the THz intensity from InN with increasing electric power (${\beta}\geq{1.3}$) prevails over sub-linear laws inherent to other processes in semiconductor structures at high electric field \cite{Shubina,Shalygin}. As a result, the THz emission with the density as high as 30 $\mu$W/cm$^{2}$ can be collected from InN epilayers. The 2-6 THz spectral range occupied by the emission band overlaps the range of the spectral sensitivity of a standard detector (2.4-5 THz).  Thus, full ability of such detectors can be exploited using an emitter with random grating.

In conclusion, our studies of InN have exhibited that the degenerate semiconductors possessing an electron accumulation layer can produce terahertz radiation related to surface plasmon polaritons. An electron temperature increase, provided by applying electric field, populates various electronic states. In the plasmon momentum distribution, equilibrium is broken down by the presence of the random grating formed by topographic defects. Thus, THz emission of appreciable intensity appears in the spectral range determined by both plasmon dispersion and statistics of grating wave vectors. We believe that similar nanostructured materials can be promising for various THz application.

We thank M. I. Dyakonov for fruitful discussions; V. N. Jmerik and A. M. Mizerov for sample growth; A. O. Zakhar'in and A. N. Sofronov for the help in THz studies; A. Yoshikawa and W. Schaff for reference samples supplying. This work has been supported in part by the RFBR and the Program of the Presidium of RAN.

\end{document}